\documentclass{aastex}
\usepackage{emulateapj5}
\usepackage{apjfonts}

	\newcommand{\citeN}[1]{\citeauthor{#1} (\citeyear{#1})}
	\newcommand{\citeNP}[1]{\citeauthor{#1} \citeyear{#1}}

	\newcommand\lineone{\rm Fe~{\sc i}~$\lambda$6301.5~\AA}
	\newcommand\linetwo{\rm Fe~{\sc i}~$\lambda$6302.5~\AA}
	\newcommand\josa{JOSA}

\shorttitle{Quiet Sun magnetic fields at high spatial resolution}
\shortauthors{Dom\'\i nguez Cerde\~na et al.}

\begin{document}
   \title{Quiet Sun magnetic fields at high spatial resolution}

   \author{I. Dom\'\i nguez Cerde\~na}
   \affil{Universit\"ats-Sternwarte,
              Geismarlandstra\ss e 11, D-37083 G\"ottingen, Germany}
   \email{ita@uni-sw.gwdg.de}

   \author{F. Kneer}
   \affil{Universit\"ats-Sternwarte,
              Geismarlandstra\ss e 11, D-37083 G\"ottingen, Germany}
   \email{kneer@uni-sw.gwdg.de}
          \and
    \author{J. S\'anchez Almeida}
    \affil{Instituto de Astrof\'\i sica de Canarias, 
              E-38200 La Laguna, Spain}
   \email{jos@ll.iac.es}

\begin{abstract}
We present spectro-polarimetric observations of Inter-Network magnetic fields 
at the solar disk center. A Fabry-Perot spectrometer 
was used to scan  the two Fe\,{\sc i} lines at $\lambda$6301.5~\AA\ and $\lambda$6302.5~\AA .
High spatial resolution  (0\farcs5) magnetograms were obtained after
speckle reconstruction. The patches with magnetic fields above noise 
cover approximately  
45\,\% of the observed area.  Such large coverage renders a mean
unsigned magnetic flux density of  some 20 G (or 20 Mx cm$^{-2}$), which exceeds 
all previous measurements.
Magnetic signals occur predominantly in intergranular spaces. 
The systematic difference between the flux densities measured in the two 
iron lines leads to the conclusion that, typically, we detect structures with
intrinsic field strengths larger than 1\,kG occupying only
2\% of the surface.
\end{abstract}
\keywords{
          Sun: magnetic fields --
          Sun: photosphere}

%
%

\section{Introduction\label{intro}}

We study magnetic fields of the quiet Sun. Our analysis will exclude the 
magnetism in photospheric network regions known to contain strong kG 
magnetic fields (e.g., \citeNP{ste94}). We will concentrate on the low flux features 
in the interior of 
the network which we will call, for brevity, IN fields (Inter-Network fields). 
Magnetic fields within the network cells were
already discovered in the seventies (\citeNP{liv75};
\citeNP{smi75}). 
During the last decade, with improved observational and diagnostic
capabilities, these IN fields have obtained increasing attention. 
They will allow to test our ideas on the magnetic field generation
by convective processes, a natural step towards  understanding the principles
of astrophysical dynamos 
(e.g. \citeNP{cat99a}; \citeNP{emo01}).
In addition, it has been found that the very quiet IN areas
harbor a sizeable fraction of the solar magnetic flux 
(see, e.g., \citeNP{ste82}; \citeNP{yi93};
\citeNP{soc02}).
Such estimates are based on lower limits to the still unknown amount 
of flux,
which suggests the potential importance of the IN component of
the solar magnetism. 

So far only a fraction of the IN field has been detected. This conclusion
follows from the presence of unresolved mixed polarities in  
1\arcsec\ angular resolution  observations
(e.g., \citeNP{san96}; \citeNP{san00}; \citeNP{lit02}). The mixing
of polarities reduces the polarization,
making the magnetic structures difficult to identify.
Consequently,
it is to be expected finding more sites with magnetic
fields when increasing the spatial resolution. This idea triggered 
our investigation: we aimed at observing the IN fields with the highest possible
spatial resolution yet keeping a good polarimetric sensitivity. 

Many of the recent observational studies find that the IN fields have 
magnetic field strengths substantially lower than 1~kG
	(\citeNP{kel94};
	\citeNP{lin95};
	\citeNP{lin99};
	\citeNP{bia98};
	\citeNP{bia99};
	\citeNP{kho02}).
Yet, on the 
basis of different observing and interpretation techniques, 
one can find in IN areas strong kG magnetic 
fields (\citeNP{gro96};
	\citeNP{sig99};
	\citeNP{san00};
	\citeNP{soc02}).
This seeming discrepancy probably
points out that the IN regions present a continuous distribution of field strengths. 
Depending on the specificities of the diagnostic technique,
one selects only a particular part of such distribution (see \citeNP{cat99a};
	\citeNP{san00}).

In this study we confirm the presence of kG IN magnetic fields. In addition,
we detect an amount of unsigned magnetic flux which exceeded any value previously 
reported in the literature. 
Our conclusions are based on spectro-polarimetric observations achieving both high spatial 
resolution (0\farcs5) and fair sensitivity (20 G). 
In this 
letter we give a short description of the observations and the data 
analysis. Our aim is to communicate the results bearing on strong IN 
fields and large flux densities.
A detailed discussion, including additional properties on the time
evolution,
will be given in a forthcoming more extended contribution. 

%
%

\section{Observations and data reduction}

The observations were obtained in April 2002 with the {\em G\"ottingen}
Fabry-Perot Interferometer
(FPI; \citeNP{ben92}; \citeNP{ben93})
mounted at the Vacuum
Tower Telescope of the Observatorio del Teide (Tenerife). The seeing was excellent 
(Fried parameter $r_0$\,=\,13--14\,cm). A very quiet region 
near disk center was selected using G band video images to avoid network 
regions. The observational setup is described in \citeN{kos01}, which includes a 
Stokes $V$ polarimeter. While scanning across a spectral 
region, the spectrometer takes two-dimensional narrow-band images with short exposure times 
and, strictly simultaneously, broadband images from the same Field Of View (FOV).

The data for the present study stem from a 17\,mininutes time series of spectral scans 
in the 6302\,\AA\ region, containing the \lineone\ line (Land\'e 
factor $g=1.67$), the \linetwo\ line ($g=2.5$), and 
the telluric O$_2$ 6302.76\,\AA\ line.
The FWHM of the FPI and the sampling interval were 44\,m\AA\ and 
32\,m\AA, respectively. Each FPI scan consists of 28 
wavelength positions for the two iron lines, and 
5 wavelength positions for the telluric line. 
The duration of a 
spectral scan is 35\,s,  plus 15s needed for storage 
onto hard disk. The FOV corresponds to 15\arcsec$\times$25\arcsec
on the Sun, sampled with a pixel size of 
0\farcs1$\times$0\farcs1.
The data reduction includes subtraction of dark current, flat fielding, and
speckle reconstruction of the broadband images 
(\citeNP{deb92};  \citeNP{deb96})
using the spectral ratio method (\citeNP{von84}) and the speckle
masking method (\citeNP{wei77}). The reconstruction of the
narrow-band images proceeds the way proposed by 
\citeauthor{kel92b} (\citeyear{kel92b}; see also \citeNP{kri99};
\citeNP{hir01}).
We use the 
code by \citeN{jan03} with
minor modifications.
To have the noise filtering independent of the seeing, which may vary during a 
spectral scan, we apply a filter which limits the spatial resolution of the 
reconstructed spectro-polarimetric images to 0\farcs5. 
To further suppress noise, the 
images  were processed with a 5$\times$5 pixel boxcar 
smoothing after the filtering. 
The FPI simultaneously collects the left and right circularly
polarized beams onto different areas of the CCD. They are added and 
subtracted to obtain the Stokes $I$ and $V$ profiles. This
task demands a careful superposition of the two beams,
which we carry out after a  
sub-pixel interpolation. 
The spatial filters were not applied to the broad-band images.

\subsection{Calibration of the magnetograms}

	Magnetic flux densities are computed starting from the 
magnetograph equation, which
relates the circular polarization $V$ and
the longitudinal component
of the magnetic field $B$ (e.g., \citeNP{unn56}; \citeNP{lan92}),
\begin{equation} 
V(\lambda) = C(\lambda)\ B, 
\label{eeqmag} 
\end{equation} 
with the calibration constant $C(\lambda)$ proportional to $dI(\lambda)/d\lambda$.
First, the
derivative of Stokes $I$ required to evaluate 
$C(\lambda)$ is computed numerically for each iron line.
Then, the wavelengths of the two extrema of the 
derivative, plus the two wavelengths immediately adjacent to 
each one of them, are selected. 
The flux density of each solar point, $B_\mathrm{eff}$, is obtained solving 
equation (\ref{eeqmag}) for these six wavelengths $\lambda_i$.
The least-squares fit renders
\begin{equation}
B_\mathrm{eff}=\sum_i V(\lambda_i)\ C(\lambda_i)\Big/\sum_i C^{2}(\lambda_i).
\label{llest}
\end{equation} 
Obviously, if 
the observed Stokes $I$ and $V$ profiles satisfy equation (\ref{eeqmag}) 
then $B_\mathrm{eff}=B$.
In general this is not the case and $B_\mathrm{eff}$ represents a mere
biased estimate of the true flux density. 

	Two main sources of noise limit the precision of $B_\mathrm{eff}$.
The measure is affected by the random noise of the 
Stokes $V$ spectra.
We estimate its influence applying to equation (\ref{llest}) the law of propagation 
of errors (e.g., \citeNP{mar71}). Assuming
the Stokes $V$ noise to be constant and equal to $0.5\%I_c$ ($I_c$ being 
the continuum intensity),
the typical value for the random error of $B_\mathrm{eff}$ turns out to be
20 G (Table \ref{table}; 1~G~$\equiv$~1~Mx~cm$^{-2}$).
The adopted $V$-noise comes from the
r.m.s. fluctuations of the Stokes $V$ profiles when $B_\mathrm{eff}\longrightarrow 0$.
A second source of error affecting $B_\mathrm{eff}$ is due to the
contamination of the polarization signals with intensity (the so-called
$I$-to-$V$ crosstalk). It can be created in many ways during the reduction
procedure (e.g., errors in setting the continuum intensity level). 
Since $V \ll I$, a small crosstalk
compromises the polarimetric accuracy of the measurements. Fortunately,
the procedure that we employ to determine $B_\mathrm{eff}$ is 
insensitive to this contamination. The calibration constant $C(\lambda)$
has different signs in the two wings of the spectral lines, consequently,
any symmetric contribution to $V$ (e.g., the contamination with 
Stokes $I$) leaves only a small residual in equation (\ref{llest}). 
Using quiet Sun Stokes $I$ profiles,
we deduce a residual of at most a few G, which is
regarded as negligible.
%
%

\section{Results and discussion}
%

\subsection{Magnetograms}

\begin{figure}
\epsscale{0.6}
\plotone{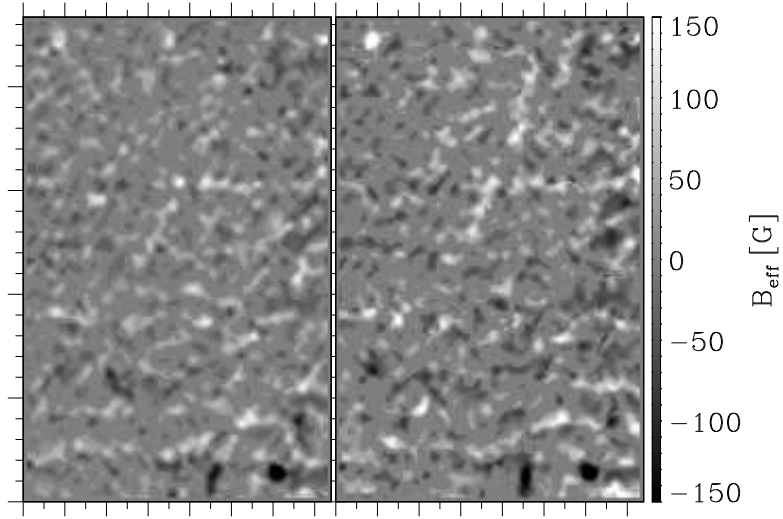}
   \caption{Magnetograms in the two Fe\,{\sc i} lines at $\lambda$6302.5\,\AA\ (left)
   and 
$\lambda$6301.5\,\AA\ (right).   
Dark and light correspond to different polarities
(see the vertical bar). The distance between
tickmarks is 1\arcsec.}
	\label{fig1}
\end{figure}

Figure \ref{fig1} presents two magnetograms taken simultaneously in the two iron 
lines. They exhibit a {\em salt and pepper} pattern. The close 
similarity of the two magnetograms demonstrates that we have found magnetic 
field signals above 
noise. We sought and found further evidences for the consistency
of the magnetograms. Among them (a) 
most of the patches have sizes of 
the resolution limit or larger, and (b)
the patches maintain their identity between successive snapshots of the
time sequence that we obtained.
Approximately 45\,\% of the FOV contains 
polarimetric signal.
The mean unsigned magnetic field density averaged over the FOV is
21~G, for the 
magnetogram based on \lineone , and 17 G for 
\linetwo\ (Table \ref{table}).
In this average we set to zero all those pixels with $B_\mathrm{eff}$
below the noise level.
The unsigned flux densities that we measure are larger than the values found 
hitherto (e.g., see the values compiled by \citeNP{san03}, \S 4.1). We detect more signals 
probably due to the unique combination of high
spatial resolution and good sensitivity of our observation.
The magnetic flux detected in the IN 
critically depends on the polarimetric
sensitivity and the angular resolution. For example,
if the magnetograms in Figure \ref{fig1} had
a sensitivity of 100~G, then the mean unsigned flux density
drops to 1-2 G (Table \ref{table};
the new lower sensitivity is modeled  setting to zero all the observed signals 
below 100 G).
The  drastic decrease of signals with decreasing sensitivity 
can easily explain why the IN signals that we detect
have been missed in previous
high resolution magnetograms 
(\citeNP{kel95}; \citeNP{sto00}; \citeNP{ber01}).
A decrease of angular resolution also reduces the signals 
due to cancellation between close 
opposite polarities.
If the angular resolution of the observation is artificially degraded 
by convolving the magnetograms in Figure \ref{fig1} with a 1\arcsec~FWHM Gaussian, 
then the mean unsigned flux densities  become of the order of 7-9 G.
These values agree with previous determinations 
based on scanning spectro-polarimeters that reach 1\arcsec\ resolution 
(\citeNP{san00} obtain 10\,G, whereas \citeNP{lit02} gives 
9\,G).  The mean {\em signed}
magnetic flux density in Figure \ref{fig1} is found to be 
+3~G for \linetwo ,  
and +2~G for \lineone.

\begin{figure}
\epsscale{.60}
\plotone{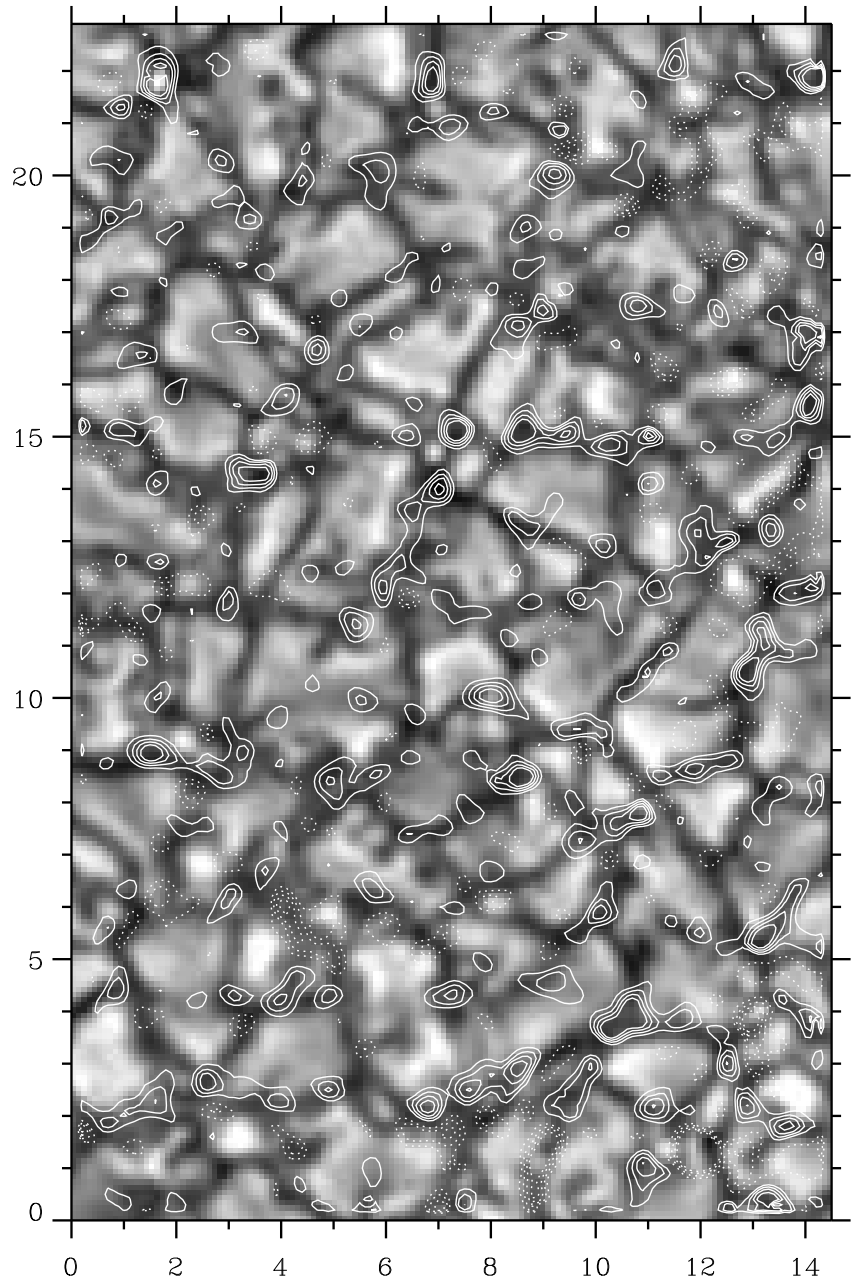}
\caption{Speckle reconstructed broadband image overlaid with the magnetogram  of
\linetwo\ with contours at $\vert B_\mathrm{eff}\vert=$~30, 50, 
70, and 90\,G. The solid and dotted contours indicate opposite polarities. The 
distance between tickmarks is 1\arcsec.}
\label{fig2}
\end{figure}

Figure \ref{fig2} shows the speckle reconstructed broad-band image 
and, overlaid to it, the 
\linetwo\ magnetogram.  
It is important to note that most of the magnetic fields, 
$\approx65$\,\%, are located in intergranular spaces. This was observed before 
(\citeNP{lin99}; \citeNP{lit02}; \citeNP{soc03b}) and is expected from 
numerical simulations of magneto-convection (e.g., \citeNP{cat99a}).
Our high spatial resolution observations 
leave no doubt on these results. However, magnetic fields can also be found in 
granules (\citeNP{sto00}; \citeNP{kos01}).
Patches of opposite polarity are often located close to each other, again in 
agreement with expectations from numerical simulations. 
%

%
\subsection{Magnetic field strengths \label{mfs}}

The polarization signals obtained from the two spectral
lines  are correlated (see Fig. \ref{fig1}). However,  
the magnitude of the effective flux density of \lineone, 
$B_\mathrm{eff}(6301)$, is systematically larger than the
effective field derived using \linetwo, $B_\mathrm{eff}(6302)$.
We have estimated
this excess by several means:
least-squares fit of $B_\mathrm{eff}(6301)$ versus $B_\mathrm{eff}(6302)$,
least-squares fit of $B_\mathrm{eff}(6302)$ versus $B_\mathrm{eff}(6301)$, 
mean ratio among all points in the field of view, etc.
Our best estimate for this ratio is,
\begin{equation}
B_\mathrm{eff}(6301) /B_\mathrm{eff}(6302)\simeq 1.28\pm 0.10,
	\label{ratio0}
\end{equation}
where the error bar accounts for all the individual estimates. 
We interpret this systematic difference 
as an indication that the magnetic field strengths in the
magnetograms are, typically, in the kG regime.
For weak fields, say 500\,G, 
the magnetograph equation (\ref{eeqmag}) holds, since
the basic conditions for the approximation to be valid are satisfied (\citeNP{soc02}).
Thus, for weak fields one expects 
$B_\mathrm{eff}(6301)/B_\mathrm{eff}(6302)\simeq1$. Since this is not the observed ratio,
one is forced to conclude 
that the field strengths have to be in the kG range.
We have modeled the ratio (\ref{ratio0})
using synthetic polarized spectra formed in atmospheres whose magnetic fields are known.
Such numerical calibration confirms the qualitative argument posed above, namely,
that the 
ratio (\ref{ratio0}) corresponds  to kG fields.

Note that structures with intrinsic kG magnetic field strength showing
20 G flux density have to
occupy only a small fraction of the solar
surface. The simplest estimate yields 2\% of the area, which
comes from
fill factor $\sim B_\mathrm{eff}/B\sim 20~{\rm G} / 1000~$G.

Magnetic structures with kG fields should be bright at the granulation level 
because they have lower density and thus lower opacity than the ambient 
atmosphere (\citeNP{spr76}). Except for a few cases (e.g.,
Fig. \ref{fig2}, upper left corner at [1\farcs5,~22\arcsec]),
we do not see such brightening in the broadband 
images. This implies that the structures must thus be smaller 
than the resolution of these images, of approximately 
0\farcs25 or 180\,km.

%
%

\section{Conclusion}

We have demonstrated that polarimetric observations with high spatial resolution 
reveal a wealth of magnetic structures even in the very quiet Sun, away from 
activity or network features. Our study reveals a mean unsigned flux density
of 20 G, which is at least a factor two larger than the flux found in previous studies
having lower angular resolution. Even more, it is almost twice the flux density
detected during the solar maximum in the form of
active regions and network using conventional techniques (see \citeNP{soc02};
	\citeNP{san02b}).
Our analysis also leads to the conclusion that the 
IN fields contain strong kG fields. 
Since only some 2\% of the 
solar surface produces these kG signals,
we do not exclude the
existence of weaker fields in the region
(of the order of tens of G, as found from Hanle diagnostics, or hundreds of G, as deduced
from Fe {\sc i} 15648~\AA\ measurements; see \S \ref{intro}).

We find no evidence that the angular resolution and sensitivity
of our magnetograms suffice to single out all the magnetic features existing in
the IN region. Consequently, the flux that we detect
should be regarded only as a lower
limit.
%
%

\acknowledgements
Thanks are due to T. Berger, H. Socas-Navarro and A. Title for discussions
on the work. 
Thanks are also due to K. Jan\ss en for letting us use her image reconstruction
routines.
IDC acknowledges support by the Deutsche Forschungsgemeinschaft (DFG) through
grant 418\,SPA-112/14/01. 
The Vacuum Tower Telescope is operated by the Kiepenheuer-Institut f\"ur
Sonnenphysik, Freiburg, at the Spanish Observatorio del Teide of the
Instituto de Astrof\'\i sica de Canarias.
The work has been partly funded by the Spanish Ministry of Science
and Technology, project AYA2001-1649.

%
%
%


%
%

%

%
%

\begin{deluxetable}{lccccc}
\tablewidth{0pt}
\tablecaption{Mean unsigned flux density in the magnetograms\label{table}}
\tablehead{line&full data& $>100$~G\tablenotemark{a}& 1\arcsec\ seeing& 
	signed flux\tablenotemark{b}& random noise\tablenotemark{c}} 
\startdata
\lineone&21 G&2 G&9 G&2 G&23 G\\
\linetwo&17 G&1 G&7 G&3 G&17 G\\
\enddata
\tablenotetext{a}{Only signals above 100 G are considered.}
\tablenotetext{b}{Contrarily to the other estimates in the table,
	the signs of the signals are considered
	in the average.}
\tablenotetext{c}{Corresponding to a single pixel.}
\end{deluxetable}

\end{document}